\documentclass[twocolumn,preprintnumbers,amsmath,amssymb,prb]{revtex4-2}
\usepackage{graphicx}
\usepackage{ulem}
\usepackage[svgnames]{xcolor}
\usepackage{bm}
\usepackage{multirow}
\usepackage{amsmath}
\usepackage{subcaption}
\newif\ifshowcomments\showcommentstrue

\begin{document}

\title{Anti-chiral order and spin reorientation transitions of
  triangle-based antiferromagnets}

\author{Leon Balents$^{1,2}$} 

\affiliation{(1)Kavli Institute for Theoretical Physics, University of California, Santa Barbara, California 93106-4030, USA \\ (2) Canadian Institute for Advanced Research, Toronto, Ontario, Canada}

\date{\today}
\begin{abstract}
We show that triangle-based antiferromagnets with ``anti-chiral''
order display a non-trivial dependence of the spin orientation with an
in-plane field.  The spins evolve from 
rotating in the opposite sense to the field at very low fields to
rotating in the same sense as the field above some critical field
scale.  In the latter regime the system displays first order
transitions at which the spin angles jump, and these first order lines
terminate in critical points in the Ising universality class.
Application to Mn$_3$Sn is discussed.
\end{abstract}
\maketitle

The elementary unit of a triangle of spins is often considered the
building block of frustrated magnetism.  Three spins on such a
triangle with antiferromagnetic Heisenberg interactions enjoy, in the classical
limit, an O(3) rotational degeneracy of ground states in which the
spins lie in a plane at 120 degree angles to one another.  When a
field is applied, the degeneracy enlarges from this symmetry-mandated
one to an accidental degeneracy which includes both coplanar and
non-coplanar states.  When such triangles are assembled into the
canonical triangular lattice, thermal and quantum fluctuations are
known to break this degeneracy in favor of the coplanar ones, a
phenomena known as ``order by disorder''\cite{starykh2015unusual}.    Larger degeneracies are
found when the triangles are more weakly connected, as in the famous
kagom\'e lattice.  There the Heisenberg degeneracy becomes extensive,
and ordering is strongly suppressed.  Commonly in real materials,
weak symmetry breaking effects such as Dzyaloshinskii-Moriya (DM) coupling
and single-ion anisotropy (SIA) provide another degeneracy breaking
mechanism leading to a selection of three-sublattice ordered states.

In this paper, we study a very common situation of
three-sublattice order based on triangles in which the Heisenberg O(3)
symmetry of the Hamiltonian is weakly broken by DM and SIA in favor of
coplanar order in zero applied field.  We adopt a symmetry-based
approach based on order parameters, which is more universal than
microscopic models of specific exchange interactions, but which
incorporates a hierarchy of coupling strengths.  In particular, we
assume that Heisenberg exchange $J$ is the largest scale, followed by
DM with strength $D$ and SIA of strength $K$, i.e. $J \gg D \gg K$.
This is inspired by the breathing kagom\'e lattice materials Mn$_3$Sn\cite{Nakatsuji2015}
and Mn$_3$Ge, but is very typical for third row transition metal
magnets.  We focus on the {\em anti-chiral} state (selected for $D>0$,
see Eq.~\eqref{eq:35}), in which, proceeding clockwise around the
triangle, spins rotate {\em counter}-clockwise (Fig.~\ref{fig:fig1}).  Spins in the
anti-chiral state are nearly free to rotate globally (see below).
\begin{figure}[h!]
  \includegraphics[width=0.8\columnwidth]{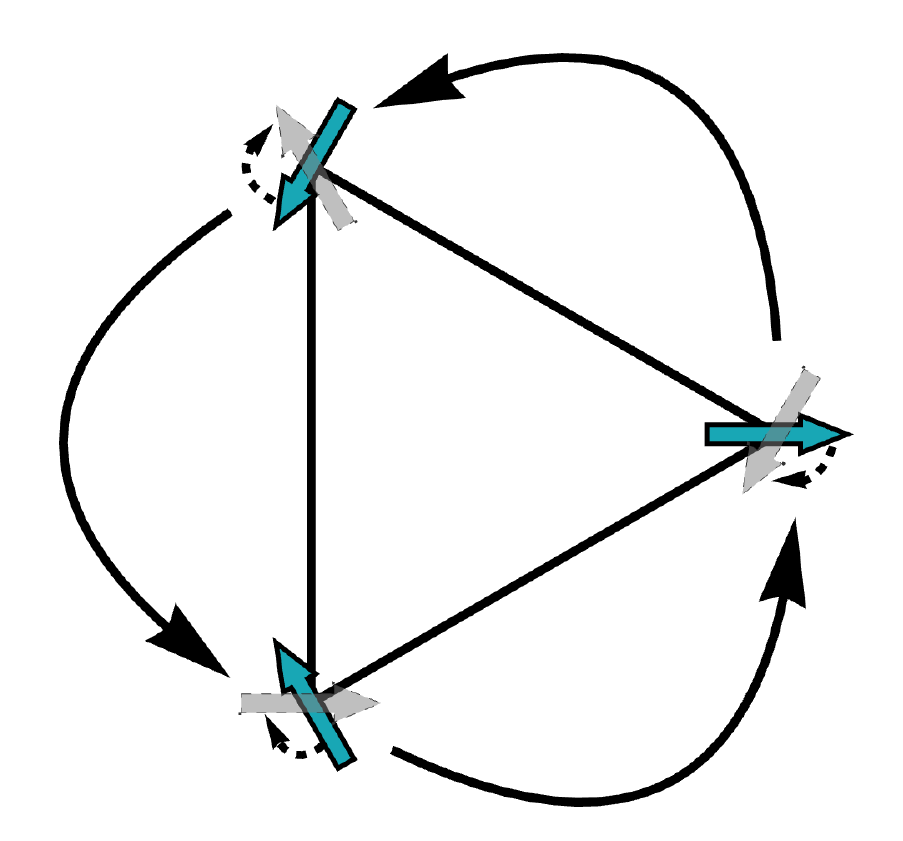}
  \caption{Illustration of the behavior of anti-chiral spins under
    rotations.  A {\em counter-clockwise} rigid rotation of the triangle by
    120 degrees, indicated by the solid arrows, rotates the spins from
    their initial state (shown in blue) to a new state, shown in gray.
    This is equivalent to a {\em clockwise} rotation of the
    spins by 120 degrees (indicated by the smaller dashed arrows) in place. }
  \label{fig:fig1}
\end{figure}
From this perspective, we consider the
evolution of the spin configurations in an applied field, and in
particular how the spins rotate when the field is rotated within the
XY plane favored by DM coupling.  We show that there is an emergent
low magnetic field scale $H^*$ separating two distinct behaviors.  When the
field is much smaller than $H^*$, the angle of a single spin within
the plane rotates in the opposite sense as the field, i.e. if the field is
oriented at an angle $\theta$ in this plane, each spin rotates with
angle $\phi_n = \phi^{(0)}_n-\theta$, where $\phi_n^{(0)}$ is an offset
for each sublattice $n$.  Conversely, when the field is much larger
than $H^*$, the spins rotate in sync with the field, i.e. $\phi_n =
\phi_n^{(0)} + \theta$.  These competing tendencies result in abrupt and
discontinuous changes in the spin configurations, which form lines of
first order transitions in the plane of the magnetic field,
terminating at second order Ising critical points (see Fig.~\ref{fig:torque}).  We argue that
features recently observed in sensitive measurements of the angular
dependence of magnetization and torque in Mn$_3$Sn are precursors of
these transitions, and that the transitions should be observable in
higher magnetic fields.

{\em Symmetry and order parameters:} We begin by presenting a
derivation of the free energy as a function of spin angle based on
symmetry and the hierarchy of energy scales.  We assume at the outset that we have a magnetic system whose ordered
structure is fully specified by giving the orientation of a set of
three spins on an elementary triangle.  We further assume that the
dominant interactions between these three spins are antiferromagnetic
and isotropic,
\begin{equation}
  \label{eq:1}
  H_0 = J \left(\bm{S}_0\cdot\bm{S}_1+\bm{S}_1\cdot\bm{S}_2+\bm{S}_2\cdot\bm{S}_0\right),
\end{equation}
with $J>0$.  This favors an ordered state in which the three spins sum
to zero, and have equal magnitudes of their expectation values.  Such
states can be written as
\begin{equation}
  \label{eq:2}
  \langle \bm{S}_n\rangle = \textrm{Re}\,\left[ \bm{d} \,e^{\frac{2\pi i n}{3}}\right],
\end{equation}
where $\bm{d}$ is a complex vector satisfying
\begin{equation}
  \label{eq:3}
  \bm{d}\cdot\bm{d} = 0.
\end{equation}
The last relation is required for the magnitude
$|\langle\bm{S}_n\rangle|$ to be independent of $n=0,1,2$.  This is
not a requirement, but would be expected at low temperature
classically.  In general, the order parameter can also be written in
terms of two orthogonal real vectors of equal magnitude, $\bm{d} =
\bm{u} + i \bm{v}$, where Eq.~\eqref{eq:3} implies
$ |\bm{u}|^2=|\bm{v}|^2$, $\bm{u}\cdot\bm{v}=0$.
These two vectors $\bm{u},\bm{v}$ define a plane in which the spins
lie.  From this one can define a third vector normal to the plane,
$\bm{w}=\bm{u}\times\bm{v} = \frac{1}{2}\textrm{Im} \left(\bm{d}^*\times \bm{d}\right)$.

In the following, we use the assumption that perturbations from the
Heisenberg limit, i.e. DM, SIA, and applied field, are all small
compared to $J$.  Then deviations from the above form are small and
more importantly they can be considered to be induced by the
perturbations.  In the effective field theory (or Landau) sense such
deviations correspond to massive modes or subdominant order
parameters, and can be integrated out order by order in the
perturbations.  This allows one to work with an effective free energy
which is a function of $\bm{d}$ only (satisfying Eq.~\eqref{eq:3}),
but in which the strength of perturbations may enter non-linearly.

We construct this free energy based on symmetry and the hierarchy of
interactions.  First, consider the symmetries in the isotropic limit
where $D=K=0$, i.e. with the Hamiltonian in Eq.~\eqref{eq:1}.  There
is in this case a global SO(3) symmetry under $\bm{S}_n \rightarrow
{\sf O} \bm{S}_n$, where ${\sf O}$ is an arbitrary SO(3) matrix.  From
Eq.~\eqref{eq:2}, this takes $\bm{d}\rightarrow {\sf O}\bm{d}$.
Second, Eq.~\eqref{eq:1} has full $S_3$ symmetry under arbitrary
permutations of the three spins.  It is convenient to regard the
permutation symmetry as generated by a $\mathbb{Z}_3$ cyclic
permutation which takes $\bm{S}_n \rightarrow \bm{S}_{n+1}$ and a
$\mathbb{Z}_2$ permutation which exchanges $\bm{S}_{1} \leftrightarrow
\bm{S}_2$.  Under these two operations, respectively, we have $\bm{d}
\rightarrow e^{2\pi i/3} \bm{d}$ and $\bm{d} \rightarrow \bm{d}^*$.

In zero magnetic field, the only non-zero invariant (using
Eq.~\eqref{eq:3}) under all these symmetries is
$\bm{d}^*\cdot \bm{d}^{\vphantom*}$, so the zero field free energy in
the isotropic limit is a function of this quantity alone.  This
dependence can be regarded as simply fixing the overall magnitude of
the order parameter, $\bm{d}^*\cdot\bm{d}^{\vphantom*} = 2n_0^2$,
where $n_0$ is the size of a local moment.  While this may shift
slightly as anisotropy and field are turned on, the effects can be
absorbed in other terms, and we can treat it, following the Landau
logic, as fixed.

With this understanding, we now introduce the magnetic field $\bm{h}$
on the isotropic spins.  It transforms in the same was $\bm{d}$ under
global SO(3) rotations, and is invariant under all the permutations.
Consequently, we find that the purely field-induced terms in the free
energy are of the form
\begin{equation}
  \label{eq:11}
  F^{\rm iso}_h =c_1 |\bm{h}\cdot \bm{d}|^2 + c_2 {\rm Re} \left[\left(\bm{h}\cdot
  \bm{d}\right)^3\right]+O(h^4).
\end{equation}
As is typical for an antiferromagnet, there is no linear coupling of
the field to the order parameter, but in this case both quadratic and
cubic terms occur\cite{PhysRevB.87.165123,PhysRevB.54.353}.

The physical meaning of these terms is as follows.  The leading
quadratic term selects configurations in which the spins lie in a
plane either normal to or containing the field, for $c_1>0$ and
$c_1<0$, respectively.  Note that the form in Eq.~\eqref{eq:2} only
defines the antiferromagnetic components of the spins (the primary
order parameter), and not the field-induced uniform moment.  For the
semi-classical Heisenberg antiferromagnet on the triangular lattice,
the two types of orderings are classically degenerate (i.e. at
$1/S=T=0$), but it is known that the coplanar configurations are
favored by both thermal and quantum fluctuations, which selects
$c_1 <0$ ($c_1 \sim - 1/(JS)$ at $T=0$)\cite{chubukov1991quantum}.
The cubic term selects an orientation of the spins within this plane:
when the sign of $c_2$ is positive (negative), one of the three spins
lies anti-parallel (parallel) to the field.
According to Ref.\onlinecite{chubukov1991quantum}, for the triangular
lattice the preferred configuration of the former type, and $c_2>0$
($c_2 \sim 1/(J^2S)$ in at $T=0$).  The same signs are found for the
classical kagom\'e lattice at non-zero temperature due to thermal
fluctuations (though the estimates differ quantitatively due to the
higher degeneracy of the kagom\'e case)\cite{gvozdikova2011magnetic}.  

Now consider the effects of DM and SIA, of the microscopic form
\begin{align}
  \label{eq:35}
  H' = \sum_n \left[D\bm{\hat{z}}\cdot \bm{S}_n \times \bm{S}_{n+1} -
  K \left(\bm{\hat{e}}_n \cdot \bm{S}_n\right)^2\right],
\end{align}
where
$\bm{\hat{e}}_n = (\cos (\frac{2\pi n}{3}),\sin (\frac{2\pi n}{3}),
0)$.  These additional terms lower the symmetry as follows.  The DM
interaction $D$ maintains a global SO(2)/U(1) subgroup of SO(3) under
rotations about the $\bm{\hat{z}}$ axis,  under the $\mathbb{Z}_3$
cyclic permutation of the spins, and under the spin-orbit coupled
$C_2$ symmetry in which the $\mathbb{Z}_2$ spin permutation discussed
earlier is combined with the corresponding rotation in spin space:
\begin{equation}
  \label{eq:12}
  C_2: \qquad \bm{S}_0 \rightarrow {\sf O}_2 \bm{S}_0 \qquad \bm{S}_{1/2}
  \rightarrow {\sf O}_2 \bm{S}_{2/1},
\end{equation}
where ${\sf O}_2 = \textrm{diag} (1,-1,-1)$.  With the SIA term $K$,
the symmetry is further reduced, so that the global SO(2) and
$\mathbb{Z}_3$ operations are collapsed to a single $C_3$ combined rotation
\begin{equation}
  \label{eq:8}
  C_3: \qquad \bm{S}_n \rightarrow {\sf O}_3 \bm{S}_{n+1},
\end{equation}
where ${\sf O}_3$ is the appropriate rotation
matrix.\footnote{Explicitly ${\sf O}_3 = \begin{pmatrix} \cos \frac{2\pi}{3} & \sin
    \frac{2\pi}{3} & 0 \\ -\sin \frac{2\pi}{3} & \cos \frac{2\pi}{3} &
    0 \\ 0 & 0 & 1\end{pmatrix}$.}

To incorporate the symmetry-lowering effects, it is convenient to
adopt a new basis
\begin{equation}
  \label{eq:14}
  d_\pm = \frac{1}{2}(d_x \pm i d_y),
\end{equation}
and trade $\bm{d}$ for $d_+, d_-$ and $d_z$.  Note that because
$\bm{d}$ is complex, $d_+$ is {\em not} the conjugate of $d_-$ and is
an independent complex field.  The symmetry operations in the new
basis become
\begin{align}
  \label{eq:15}
  SO(2): & \, d_+ \rightarrow e^{i\vartheta} d_+, && d_- \rightarrow
                                                   e^{-i\vartheta} d_-
                                                     , && d_z
                                                          \rightarrow
                                                          d_z,
                                                          \nonumber \\
  \mathbb{Z}_3: & \, d_+ \rightarrow e^{2\pi i/3} d_+, && d_- \rightarrow e^{2\pi i/3} d_- ,&&
                                                                     d_z
                                                                     \rightarrow
                                                                                              e^{2\pi i/3}   d_z,\nonumber\\
  \mathbb{Z}_2: & \, d_+ \rightarrow d_-^*, && d_- \rightarrow d_+^*,
                                                       && d_z
                                                          \rightarrow d_z^*,\nonumber\\
  C_3: & \, d_+ \rightarrow e^{4\pi i/3} d_+, && d_- \rightarrow d_- ,&&
                                                                     d_z
                                                                     \rightarrow
                                                                     e^{2\pi
                                                                     i/3}d_z,\nonumber
  \\
  C_2: &\, d_+ \rightarrow d_+^* ,&& d_- \rightarrow d_-^*, && d_z
                                                            \rightarrow
                                                            - d_z^*,\nonumber
  \\
  \mathcal{T}: & \,d_+ \rightarrow - d_+, && d_-\rightarrow -d_- ,&& d_z
                                                                  \rightarrow -d_z.
\end{align}
To summarize, the DM term is invariant under $SO(2)$, $\mathbb{Z}_3$,
$C_2$ and $\mathcal{T}$.    The $K$ term is invariant under $C_3$,$C_2$
and $\mathcal{T}$.   So the first three lines above are approximate
symmetries while the final three are exact. It is also useful to note that under
$\mathbb{Z}_2$,  $D\rightarrow -D$ (but $\mathbb{Z}_2$ does not act
simply upon $K$).

Using the above symmetries, and using the constraint Eq.~\eqref{eq:3}
and the condition that $\bm{d}^*\cdot \bm{d}=2n_0^2$, the most
general, non-constant free energy terms at zero field and quadratic in
$\bm{d}$ are
\begin{align}
  \label{eq:18}
  F_2 = &  s_1 \left(d_{+}^* d^{\vphantom*}_{+} -d_{-}^*
          d^{\vphantom*}_{-}\right)\nonumber \\
  & + s_2  \left(d_z^*d_z-2 d_{+}^* d^{\vphantom*}_{+} -2 d_{-}^*
          d^{\vphantom*}_{-} \right) + s_3 \textrm{Re} \left( d_{-}^2 \right).
\end{align}
A na\"ive calculation, simply inserting Eq.~\eqref{eq:2} in Eq.~\eqref{eq:35}, shows that $s_1 \sim -D$, $s_3 \sim -K$, while
$s_2 \sim K$, while an additional contribution $\Delta s_2 \sim D^2/J$
is expected to arise at second order in the DM coupling.  In both
cases $s_2,s_3 \ll |s_1|$.  

Here we are interested in $D>0$ which implies $s_1<0$ which favors
$d_-=d_z=0$ and $|d_+| = n_0$ (check factors).   This is the
anti-chiral state.  Then the $s_2$ term is constant and the $s_3$ term
vanishes.  Note that the phase of $d_+$ is arbitrary at this level,
reflecting the fact that the rotation of the spins is in the opposite
sense to the rotation of the local easy axes, so that the two are
incompatible.  If by contrast we take $D<0$, the chiral state with
$d_+=d_z=0$ is stabilized and $|d_-|=n_0$.  Then the $s_3$ term is
non-zero and in fact fixes the phase of $d_-$, which means the spins
are not free to rotate in the chiral state.

For the anti-chiral state, the complete freedom to rotate the phase is
an artifact of the truncation of Eq.~\eqref{eq:18} to second order in
$\bm{d}$.  A non-trivial invariant fixing the phase of $d_+$ arises
at sixth order:
\begin{equation}
  \label{eq:19}
  f_6 = \lambda \textrm{Re}\, \left( d_+^6 \right).
\end{equation}
We expect that $\lambda \sim K^3/J$, as was verified by calculations
for Mn$_3$Sn, and therefore is extremely small and often negligible.

Now consider the terms involving the magnetic field.  Similarly to
Eq.~\eqref{eq:14}, define
\begin{equation}
  \label{eq:20}
  h_\pm = h_x \pm i h_y.
\end{equation}
Note that $h_\pm^* = h_\mp$ (unlike for $d_\pm$), so it is sufficient
to list the properties of $h_+$ and $h_z$.
Under the various transformations, we have
\begin{align}
  \label{eq:21}
  C_3: & \, h_+ \rightarrow e^{2\pi i/3} h_+, && 
                                                                       h_z
                                                                       \rightarrow
                                                                       h_z,
  \\
  C_2: & \, h_+ \rightarrow h_+^*=h_-, && h_z \rightarrow - h_z, \\
  \mathcal{T}: & \, h_+ \rightarrow -h_+, && h_z \rightarrow -h_z.
\end{align}
Comparing now Eq.~\eqref{eq:15} and Eq.~\eqref{eq:21}, we can find
invariants involving the field and $d_\mu$.  To linear order in the
field, we find
\begin{align}
  \label{eq:22}
  f_{h,1} = g_1\textrm{Re}\left( h_+ d_+ \right) + g_2 \,h_z \textrm{Im}
  \left( d_-\right).
\end{align}
In Mn$_3$Sn, where the order is anti-chiral, $d_-=0$ and only the
$g_1$ term is active.  It is order of $g_1 \sim K/J$.  We see that the linear coupling to the field
multiplies $h_+$ and $d_+$, which favors rotating these complex
numbers {\em with opposite phases}.  This expresses the surprising phenomena
that each spin in the anti-chiral case at small fields actually
rotates in the opposite sense as the applied field!  The $g_1$ term
can be understood physically from the picture in Fig.~\ref{fig:fig1}:
the net effect of a rigid counter-clockwise $C_3$ rotation of the anti-chiral
triangle is the same as rotating the spins in place by a
clockwise $C_3$ rotation.  

Note that this effect contradicts the behavior in the isotropic
system, which is dictated by the cubic coupling in Eq.~\eqref{eq:11},
and favors rotating each spin in sync with the field.  The opposite
tendencies lead to a transition as a function of field strength.

To unveil it more cleanly, we focus now on the case in which only
$d_+$ is assumed non-zero, and the magnetic field is in the plane, and
write the free energy as a series in $d_+$ and the field only.  We
furthermore assume that the higher order terms in field are dominated
by the ones already present in Eq.~\eqref{eq:11}, and simply express
those in the case where $d_z=d_-=0$ in terms of $d_+$.  We find in
this case $\bm{h}\cdot\bm{d} = 1/2 h_- d_+$ which leads to
\begin{align}
  \label{eq:23}
  f_+ = &\lambda \,\textrm{Re}\, \left( d_+^6 \right) +
  g_1\textrm{Re}\left( h_+ d_+ \right) \nonumber \\
  & + \frac{c_1}{4} |h_+|^2 |d_+|^2 + \frac{c_2}{8} \textrm{Re}\left((h_-d_+)^3 \right).
\end{align}
There are many more symmetry allowed terms, but the above minimal
expression is sufficient to expose the physics and indeed one can also
show for the case of Mn$_3$Sn that all remaining terms which arise are
parametrically small in the regime of interest when $K\ll J$.

To analyze Eq.~\eqref{eq:23}, we change to angular coordinates, $h_+ =
h e^{i\theta}$ and $d_+=d e^{i\phi}$.  It becomes, up to a constant,
\begin{align}
  \label{eq:24}
  f_+ =   -w \cos 6\phi  
       - u h \cos (\phi+\theta)- v h^3 \cos 3(\phi-\theta),
\end{align}
where
\begin{equation}
  \label{eq:25}
  w = -\lambda d^6, \qquad u = -g_1 d, \qquad v = -c_2 d^3/8.
\end{equation}
{\em Angular analysis and phase transitions:} Eq.~\eqref{eq:24} is the
general result for the angle-dependent free energy of the anti-chiral
state.  We now show that it exhibits the phase transitions described
in the introduction.
\begin{figure}[h!]
  \centering
  \includegraphics[width=\columnwidth]{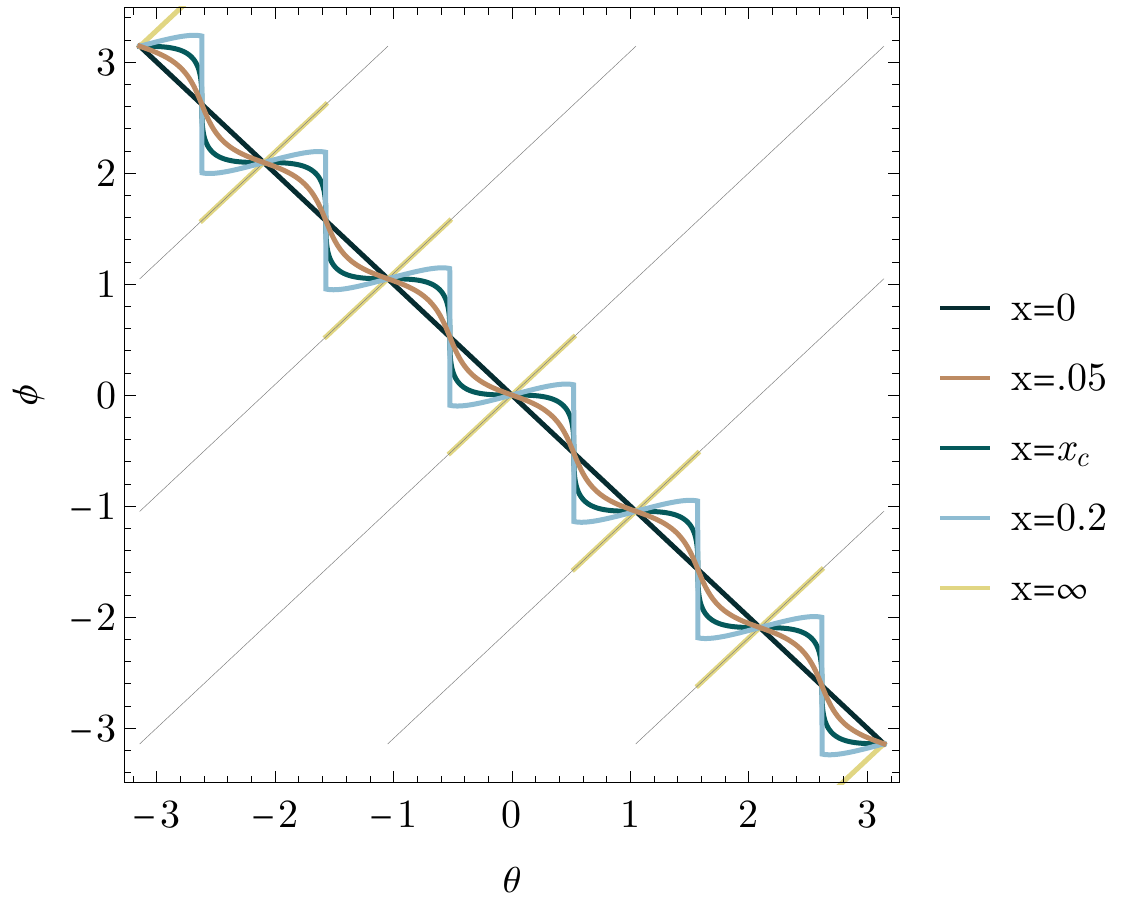}
  \caption{Spin angle $\phi$ versus field angle $\theta$ for various
    values of $x$.  The thin gray lines show the 3 degenerate branches
  at $x=\infty$, while piecewise linear function labeled $x=\infty$ is
the one selected at asymptotically large but finite $x$.}
  \label{fig:thetaphi}
\end{figure}
Without loss of generality, we take $u,v,w>0$.  Using the aforementioned estimates $w \sim K^3/J^2$,$u \sim K/J$ and
$v\sim D/J^3$, we establish the condition $w \ll
\sqrt{u^3/v}$, under which $w$ can be neglected in the field
regime $h \gg w/u$.  We henceforth assume this condition and take
$w=0$.  Then the order
parameter angle $\phi$ is determined just by minimizing the final two
terms in Eq.~\eqref{eq:24}.  So we may write $f_+ =  u h
g(\phi,\theta)$, with 
\begin{equation}
  \label{eq:26}
 g(\phi,\theta) =  -\cos (\phi+\theta) - x  \cos 3(\phi-\theta),
\end{equation}
where $x=\sqrt{v/u} h^2>0$.  The optimal spin angle $\phi(\theta,x)$
is determined from minimizing $g$ at fixed field angle $\theta$ and
$x$.  It is instructive to analyze the two limits $x=0$ and
$x=\infty$.  At $x=0$, $g$ is clearly minimized by $\phi=-\theta$.  At
$x=\infty$, there are three degenerate mnima with
$\phi = \theta + 2\pi k/3$, with $k=0,1,2$.  One observes that the
spin angles winds in the opposite sense in the two extreme limits.
The degeneracy in the large $x$ limit is resolved by selecting the
branch ($k$) which minimizes the first term in Eq.~\eqref{eq:26}.
This leads to jumps in $k$ (and hence $\phi$) as a function of
$\theta$, which occur when $\theta = \pi/6 + \pi m/3$, with integer
$m$, as shown in Fig.~\ref{fig:thetaphi}.  Away from the two extreme limits,  the curve
$\phi(\theta)$ evolves, but the discontinuities persist for large $x$,
while they are absence for small $x$.   A transition occurs for
$x=x_c$, where the the discontinuities first appear.

To clarify the critical points, we define $\psi = \phi+\theta$, and
let $\theta=\pi/6+\delta$, so that $\delta=0$ defines the location of
one of the discontinuities for $x>x_c$.  Some algebra gives
\begin{align}
   \label{eq:30}
  \tilde{g}(\psi,\delta) & = g(\psi-\tfrac{\pi}{6}-\delta,\theta) \\
  & = -\cos \psi + x\cos 6\delta
     \cos 3\psi +x \sin 6\delta \sin 3\psi.\nonumber
\end{align}
We see that at $\delta=0$, $\tilde{g}(\psi,0)$ is an even function of
$\psi$.  In fact, the even-ness of this function reflects a symmetry
$\phi \rightarrow \pi/3-\phi$, which is a $C_2$ symmetry of the
Hamiltonian when the field angle $\theta=\pi/6$.  Hence
$\tilde{g}(\psi,\delta)$ can be regarded as a Landau function, with a
minima at $\psi=0$ for $x<x_c = 1/9$, which bifurcates for $x>x_c$
into two degenerate minima at $\psi=\pm \psi_0$.  This is an Ising
phase transition.  Note that for Mn$_3$Sn, the Ising symmetry
  along these lines can be traced to a symmetry under a $C_2$ rotation
  about the axis
  of the magnetic field, when the field is aligned to these
  crystalline axes.  The deviation $\delta$ of
the field angle plays the role of a symmetry-breaking field on the
Ising order parameter, and the discontinuities in $\theta$ are
analogous to the first order transition that occurs within the ordered
phase of the Ising model on changing the sign of the field.
\begin{figure}[h!]
  \centering
  \includegraphics[width=\columnwidth]{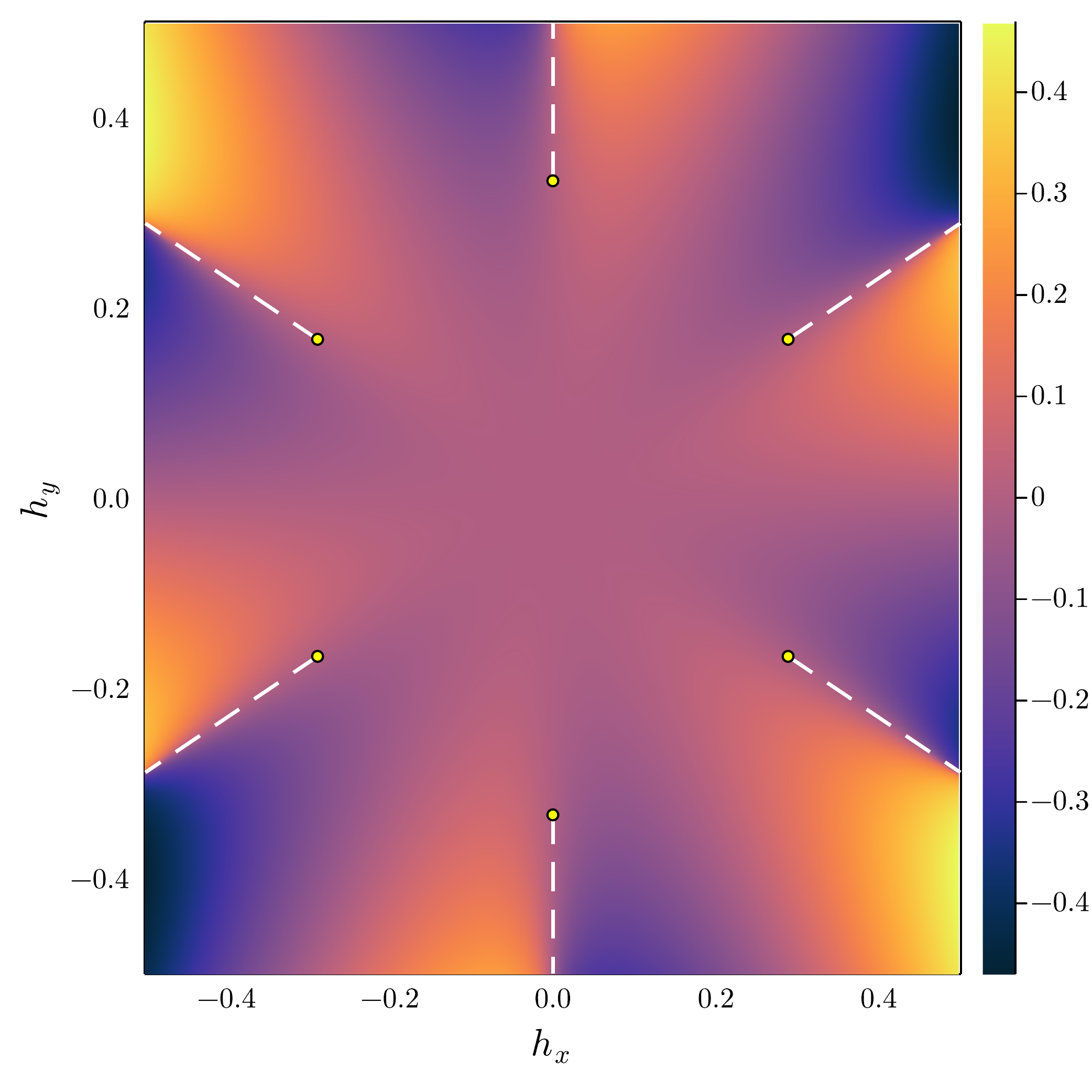}
  \caption{Density plot of torque ($\tau/u$) versus in-plane magnetic field.  The
    field scale is chosen so that $h_c = 1/3$.  Yellow circles mark
    the location of Ising critical points, and the first order lines
    are marked with dashed lines.  }
  \label{fig:torque}
\end{figure}
This analysis determines a critical field
$h_c = \frac{1}{3}\sqrt{\frac{u}{v}}$.  Following the analogy with the
Ising model, one observes within this mean field picture that at the
critical field, the angle
$\psi \sim - |\omega|^{1/3} \textrm{sign}(\omega)$ for small
variations of the field angle near $\pi/6$.  This is analogous to the
non-linear susceptibility of the Ising ferromagnet at criticality.  A
priori thermal fluctuations will renormalize this exponent to that of
the 3d Ising model -- there we have $\psi \sim |\omega|^{1/\delta}$
and the true value is approximately $\delta \approx 4.8$.

One can readily calculate the torque $\tau=df/d\theta$, where the
{\em total} derivative should be taken with respect to the field angle
of the energy optimized at $\phi=\phi(\theta)$.  Using the energy
minimization condition, one obtains $\tau = 2 uh \sin
(\phi(\theta)+\theta)$, which is plotted in Fig.~\ref{fig:torque}.  It is discontinuous across the first order
transition lines, and vanishes when $\theta$ is a multiple of
$\pi/3$.  

The angular transition may also be detectable by means other than the
torque.  Hysteresis may occur and domains may form near the first
order lines.  The spin angles themselves could be observed directly in
neutron scattering.  It is interesting to point out that a very
similar phenomena has already been observed in CeAlGe, in which the
spontaneous formation of domains was argued to give rise to a sharp
peak in the resistance versus angle, dubbed singular angular
magneto-resistance\cite{Suzuki377}.  It would be very interesting to
study the angular dependence of the resistance in Mn$_3$Sn in an
appropriate range.

From the calculations in Ref.\cite{Liu2017}, and the Supplemental
Material of Ref.\cite{li2021free}, we
can compare directly to Eq.~\eqref{eq:24}, and extract the parameters
of the symmetry based theory in terms of microscopics.   In the
classical zero temperature model, one obtains thereby
\begin{align}
  \label{eq:31}
  u = \frac{K}{J+\sqrt{3}D}, && v = \frac{D}{3\sqrt{3}(J+\sqrt{3}D)^3}.
\end{align}
This leads to the critical field, restoring units
\begin{align}
  \label{eq:32}
  H_c = \frac{J+\sqrt{3}D}{g\mu_B} \sqrt{\frac{K}{D}}.
\end{align}
Taking $D=0.2J$, $K=.006J$, $J=20meV$ and $g=3$ yields $H_c\approx
20T$.  This is of course to be renormalized by thermal and quantum
fluctuations, but gives an idea of the order of magnitude.  It
strongly suggests the transition should be within range of current
experiments.  

\begin{acknowledgments} I thank Kamran Behnia, Zengwei Zhu, and Xiaokang Li
for the experimental collaboration that inspired this work.  This
research is supported by the NSF CMMT program under Grant
No. DMR-2116515.
\end{acknowledgments}

\vfill

\bibliography{antichiral}

\end{document}